\documentclass[smallextended]{svjour3}       
\smartqed  
\usepackage{graphicx}
\usepackage{mathptmx}      
\usepackage{braket} 
\usepackage{extarrows} 
\usepackage[justification=centering]{caption} 
\usepackage{amsmath}
\usepackage{bm}
 \journalname{}
\begin{document}

\title{Quantum Differential Cryptanalysis
}


\author{Qing Zhou         \and
        Songfeng Lu \and
        Zhigang Zhang \and
        Jie Sun 
}


\institute{Qing Zhou \at
              School of Computer Science and Technology, Huazhong University of Science and Technology, Wuhan 430074, China \\
           \and
           Songfeng Lu \at
              School of Computer Science and Technology, Huazhong University of Science and Technology, Wuhan 430074, China \\
            Corresponding author,   \email{lusongfeng@hotmail.com}           
           \and
           Zhigang Zhang \at
              School of Computer Science and Technology, Huazhong University of Science and Technology, Wuhan 430074, China \\
           \and
           Jie Sun \at
              School of Computer Science and Technology, Huazhong University of Science and Technology, Wuhan 430074, China \\
}

\date{Received: date / Accepted: date}

\maketitle

\begin{abstract}
In this paper, we propose a quantum version of the differential cryptanalysis which offers a quadratic speedup over the existing classical one and show the
quantum circuit implementing it. The quantum differential cryptanalysis is based on
the quantum minimum/maximum-finding algorithm, where the values to be compared and filtered are obtained by calling the quantum counting algorithm. Any cipher which is vulnerable to the classical differential cryptanalysis based on counting
procedures can be cracked more quickly under this quantum differential attack.
\end{abstract}

\section{Introduction}
\label{intro}
Quantum computing can be used to solve many problems more efficiently than classical algorithms, including attacking classical cryptosystems. The first and most remarkable example is Shor's quantum factoring and discrete logarithm algorithms~\cite{Ref1}, which can be employed to break cryptosystems based on this two kinds of problems such as RSA and elliptic curve cryptosystems in polynomial time with high probability~\cite{Ref2}. Another significant and simple quantum approach to analyzing classical ciphers is to exhaustively search the key space with Grover's algorithm ~\cite{Ref3}, with which we are able to obtain the system key among $N$ candidates by only $O\left(\sqrt{N}\right)$ steps. Besides, Grover's algorithm can be combined with several existing classical cryptanalytic methods to attack certain types of ciphers~\cite{Ref5,Ref6,Ref7}. However, the quantum mechanical version of a general yet advanced and efficient classical attack has never been presented in the open literature.

Here, we implemented the classical differential cryptanalysis with quantum circuits and achieves a quadratic speed-up over it.
\section{Classical differential cryptanalysis}
\label{sec:2}
Differential cryptanalysis~\cite{Ref9} is one of the most important classical cryptanalysis approaches since its publication. It works on many pairs of plaintext with the same specific difference and their corresponding ciphertexts and analyzes the effect of the plaintext difference on the resultant ciphertexts differences. Although some modern block ciphers such as AES are not vulnerable to this technique, the quantum implementation of it could be helpful in the work of quantum-fying more sophisticated and mature classical attacks, and the interest of this paper is more about the study of the quantum version compared to the classical one.

Briefly speaking, the main idea of the classical differential cryptanalysis is trying to attain the system key by analyzing the evolution of the difference values throughout the encryption based on a characteristic\footnote{A characteristic is a structure built for the cipher to be attacked, which is constructed by cryptanalysts at their leisure time. An n-round characteristic associated with a pair of encryptions consists of the plaintext difference and the ciphertext difference of the pair, along with the input and output differences of each round. A characteristic has a probability, which is the chance that a random pair with the chosen plaintext difference has the round and ciphertext difference values specified by the characteristic when random independent keys are used.} constructed in advance, which usually proceeds as below:

\begin{enumerate}
\item[(1)] Calculate the expression $E$ for the expected output difference of the cipher's last round according to the characteristic. With this expression, we can decide whether a given pair is the right pair of a specific candidate subkey.\footnote{A subkey in the differential cryptanalysis refers to a part of the complete key, and it is a subset of the binary key string. For the 16-round characteristic introduced in \cite{Ref14}, $E$ is $H'=T'_L\oplus 19\,60\,00\,00_x$, where $H'$ is the output XOR of the 16th round of DES and $T'_L$ is the left half of the ciphertexts difference. On the one hand, we can compute $H'$ according to $E$ with the given pair alone; on the other hand, we can also calculate $H'$ with the help of the candidate subkey of the last round by a trial encryption. If these two results coincide perfectly, then the given pair is the right pair of the candidate subkey.}

\item[(2)] Perform the counting procedure for each of the candidate subkeys of the last round, the one who has most right pairs is regarded as the correct subkey.
\end{enumerate}

After the preceding steps, use some other characteristics to reveal the remaining subkey bits with the similar method or exhaustively search the remaining subkeys if the number of unknown bits of the complete key is small enough.

To summarize, the classical differential cryptanalysis includes two stages: 1. Count the number of the right pairs of each candidate subkey; 2. find the largest one among the counting results, which corresponds to the correct subkey. Fortunately, these two algorithms can be more efficiently accomplished with quantum counting ~\cite{Ref10} and the quantum algorithm for finding the minimum proposed by Christoph Durr and Peter H\o yer ~\cite{Ref11}.
\section{Quantum differential cryptanalysis}
\label{sec:3}
To make our description more intelligible, we explain here some frequently used variables in this and the following sections.
\begin{description}
\item[\textbf{\textit{N,~n}}:] $N$ is the number of pairs needed to be analyzed and counted, usually $N$ is a power of 2 and thus $N=2^n$;
\item[\textbf{\textit{K,~k}}:] $k$ is the number of subkey bits to be recovered and $K\equiv 2^k$;
\item[\textbf{\textit{m}}:] $m$ denotes the number of bits of accuracy when estimating the angle of rotation $\theta$ determined by the Grover iteration $G$;
\item[\bm{$\epsilon$}:] $\epsilon$ is the lower bound of the success probability of estimating $\theta$ to $m$ bits of accuracy in the quantum counting procedure;
\item[\textbf{\textit{t}}:] $t$ is the length of the binary counting results and $t\equiv m+\left\lceil \log \left(2+1/{2\epsilon}\right)\right\rceil$ ~\cite{Ref12};
\item[\textbf{\textit{M}}:] $M$ refers to the estimated number of right pairs of a specific candidate subkey.
\end{description}

According to~\cite{Ref9}, the differential analytic procedure is performed for several times, and each time a certain part of the key(i.e., a subkey) can be recovered. The time complexity of the whole attack depends on the first performance, for it reveals more key bits than the subsequent ones and the computational complexity of each performance is $\Omega\left(2^k+N\right)$, where $N$ keeps unchanged. Therefore, we just give the quantum version of the first performance, and the later quantum differential analytic process for the remaining part of the key can be carried out accordingly.
\subsection{The overall circuit}
\label{sec:3.1}
Generally speaking, the quantum differential cryptanalysis utilizes the quantum maximum-finding algorithm~\cite{Ref11} as its framework and takes the modified quantum counting algorithm as a subroutine to fulfill the differential attack. It calls the quantum counting and the quantum searching algorithms~\cite{Ref10} to find a subkey whose number of right pairs is larger than that of a specific threshold subkey, and then the result is chosen as the new threshold subkey. This process is repeated until the probability that the threshold subkey has most right pairs is sufficiently large. The quantum differential attack can be carried out with the following five steps.
\begin{enumerate}
\item[(1)] Calculate the differential expression $E$ for the expected output difference value of the cipher's last round according to the characteristic.
\item[(2)] Suppose $N=2^n$ pairs need to be analyzed to recover $k=\log _2K$ bits of the last round subkey and the plaintexts difference of the characteristic is $P'$ , then request to the cryptosystem the ciphertexts of $N$ plaintexts pairs whose difference is $P'$.\footnote{When the difference operation is XOR, the plaintext pairs can be set to $\left(P_i,P_i\oplus P'\right)$ as default, where $P_i=i, i=1,2,\ldots ,N$, thus we can just send $P'$ as the request parameter to the cryptosystem and the cryptosystem could set up the plaintexts pairs by itself, and then encrypt them, pack the result and respond.}
\item[(3)] Randomly choose a threshold subkey value $y$ from the integer set \{0,\dots,$K-1$\}.
\item[(4)] Repeat the following procedures until the total running time is more than $2cm_0$, where $m_0$ is the expected total time taken by step (4) before $y$ holds the candidate subkey who has most right pairs and $c$ is a constant number, and then go the stage (e). The magnitude of $c$ is closely related with the success probability of this maximum-finding process. The steps indicated by (a) through (e) are illustrated in Fig.~\ref{fig:1}.
\begin{figure}
  \centering \includegraphics[width=80mm]{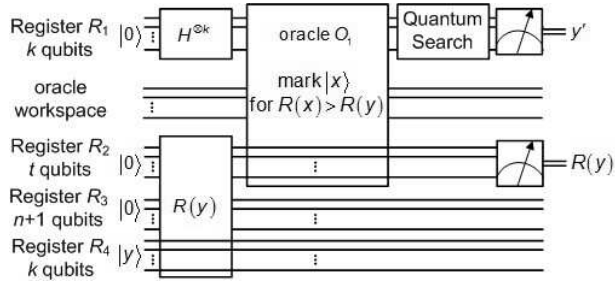}
    \caption{Schematic circuit for the quantum differential cryptanalysis}
    \label{fig:1}       
\end{figure}
    \begin{enumerate}
    \item[(a)] Initialize the registers $R_1$, $R_2$, $R_3$, $R_4$ as $\Ket{\Psi}$, $\Ket{0}^{\otimes t}$, $\Ket{0}^{\otimes (n+1)}$ and $\Ket{y}$ respectively, where $\Ket{\Psi}$ is a $k$-qubit equal superposition state:
        \begin{equation}
        \Ket{\Psi}=\frac{1}{\sqrt{K}} \sum_{i=0}^{K-1} \Ket{i},
        \label{equa:1}
        \end{equation}
    and it can be regarded as the superposition of $K$ candidate subkeys.
    \item[(b)] Let $R(y)$ be the number of right pairs of a candidate subkey whose value is $y$ (for simplicity, we will refer it shortly by 'candidate subkey $y'$ hereafter), compute $R(y)$ by performing the modified quantum counting procedure on $R_2$, $R_3$ and $R_4$. The circuit for calculating $R(y)$ will be described in the next subsection.
    \item[(c)] Apply an oracle $O_1$ onto $R_1$, which have the ability to recognize and mark the candidate subkey states in $\Ket{\Psi}$ whose number of right pairs is more than $R(y)$ with the help of the outcome of $R_2$. Let $\Ket{x}$ be a computational basis corresponding to the subkey $x$, and the action of $O_1$ can be defined by
        \begin{equation}
            \Ket{x}\Ket{q}\xlongrightarrow[]{O_1} \Ket{x} \Ket{q \oplus f(x,y)},
            \label{equa:2}
        \end{equation}
        where $f(x,y)$ equals 0 if $R(x)\leq R(y)$ and 1 if $R(x)>R(y)$ and $\Ket{q}$ is a oracle qubit. We can check whether $R(x)>R(y)$ by seeing whether $\Ket{q}$ has been flipped. With oracle $O_1$, we can mark all the items in $\Ket{\Psi}$ corresponding to subkeys who have right pairs more than $R(y)$ by only one call of $R(\cdot)$.
    \item[(d)] Apply the quantum searching algorithm on $R_1$ to get a marked item.
    \item[(e)] Observe $R_1$ and $R_2$. Let $y'$ be the outcome of $R_1$ and $R(y)$ the result of $R_2$, compute $R(y')$ and set the threshold $y$ to $y'$ if $R(y')>R(y)$.
    \end{enumerate}
\item[(5)] Return $y$ as the correct subkey value.
\end{enumerate}
\subsection{The circuit for computing $R(x)$}
\label{sec:3.2}
\begin{figure}
    \centering \includegraphics[width=90mm]{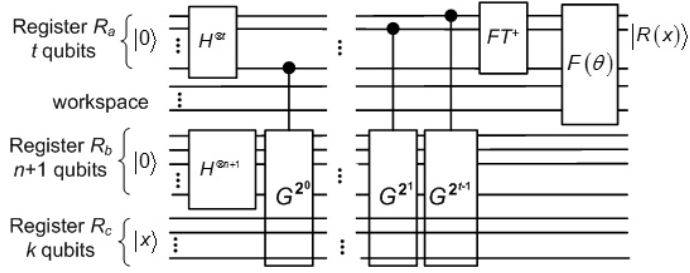}
    \caption{Circuit for calculating $R(x)$}
    \label{fig:2}
\end{figure}
The circuit for counting the number of right pairs of subkey $x$ shown in Fig.~\ref{fig:2} is generally the same as the existing quantum counting circuits \cite{Ref12} except that an additional register $R_c$ holding $x$ is included, and each Grover iteration $G$ is augmented --- the oracle in which takes $x$ as a parameter to mark the states corresponding to the right pairs of subkey $x$.

The other two registers in the new slightly modified quantum counting circuit are related to the counting result and objects: $R_a$ contains $t$ qubits and will hold the approximate counting result; $R_b$ contains $n+1$ qubits, which can be think of as a superposition of $2N$ pairs. The gate $FT^\dag$ denotes the inverse Fourier transform. If the result of $R_a$ after $FT^\dag$ is $\Ket{\theta}$, then a quantum gate computing $F(\theta)=2N\sin^2(\theta /2)$ will complete the counting process, making the final result in $R_a$ be $\Ket{R(x)}$.

\begin{figure}
    \centering \includegraphics[width=75mm]{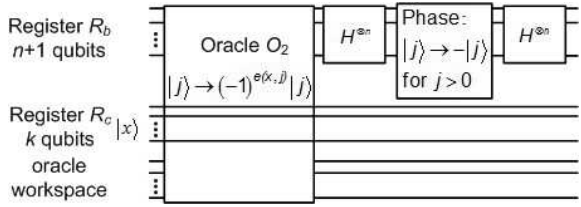}
    \caption{Circuit for the Grover iteration $G$}
    \label{fig:3}
\end{figure}

The quantum circuit of the Grover iterations $G$ in this counting procedure is depicted in Fig.~\ref{fig:3}.
$G$ takes $k+n+1$ qubits as input but only acts on the $n+1$ qubits in $R_b$, the $k$-qubit value in $R_c$ is read only. Suppose the input state to $G$ from $R_c$ is  $\Ket{x}$, the value in $R_b$ after the Hadamard transform is the equal superposition of $\Ket{j}$, $0\leq j\leq 2N-1$, then the function of oracle $O_2$ called by $G$ could be expressed as
\begin{equation}
    \Ket{j}\xlongrightarrow[]{O_2} (-1)^{e(x,j)}\Ket{j},
\end{equation}
and $e(x,j)$ can defined as
$$e(x,j)=
\begin{cases}
1& \text{if} \ Pair(j) \ \text{is a right pair of} \ x\\
0& \text{if} \ Pair(j) \ \text{cannot be a right pair of} \ x
\end{cases}$$
where $Pair(j)$ is the pair indexed by $j$. Function $e(x,j)$ can be implemented by a one round partial encryption: If the difference of the encryption results of $Pair(j)$ by $x$ equals the expected output difference calculated by $E$, then $Pair(j)$ is a right pair of $x$, otherwise $Pair(j)$ cannot be a right pair of $x$.
\section{Analysis and performance}
\label{sec:4}
\subsection{Computational complexity}
\label{sec:4.1}
For a given encryption scheme, the first two steps of the quantum differential cryptanalysis can be accomplished ahead of the attack and the pairs setup as well as the encryptions of them in step (3) are proceeded by the cryptosystem rather than the cryptanalyst, so these three steps will not be taken into account here.

By convention, we assume that any initialization on $n$ qubits takes $n$ time steps, including the Hadamard transform $H^{\otimes n}$, and that one iteration in the quantum search algorithm takes one time step. For convenience, we use $\log(x)$ to denote logarithms to base 2 and use TC$_d$ to denote the time complexity of stage $d$.

The random value in step (4) can be obtained by measuring on an equal superposition state like $\Ket{\Psi}$, which is produced by performing $k$ Hadamard gates on $\Ket{0}^{\otimes k}$, thus $\textrm{TC}_{(4)}=O(k)$.

In step 5(a), the initialization of $R_1$, $R_2$, $R_3$ and $R_4$ needs $2k+t+n+1$ time steps, which may proceeds for several times in the loop. The time it takes throughout the loop is
\begin{equation}
        \textrm{TC}_{5\textrm{(a)}}=O\left[\sum_{r=2}^{K}\frac{1}{r}(2k+t+n+1) \right]=O\left(k^2\right)+O(kt)+O(kn).
        \label{equa:4}
\end{equation}
The readers may refer to \cite{Ref11} for the detailed deduction of it.

When computing $R(y)$ in step 5(b), $t+n+1$ Hadamard transforms are used to initialize $R_a$ and $R_b$ and the number of $G$ gates in quantum counting is $2^t-1$. Besides, at most $t(t+1)/2+t/2$ gates are required for the inverse quantum Fourier transform; thus, the time complexity of 5(b) throughout the loop is
\begin{equation}
        \textrm{TC}_{5\textrm{(b)}}=O\left[\sum_{r=2}^{K}\frac{1}{r}\left(2^t+t^2/2+2t+n\right)\right]=O\left(2^tk\right)+O\left(kn\right).
        \label{equa:5}
\end{equation}

In 5(c), the oracle $O_1$ merely needs only one call of $R(\cdot)$ to mark the subkey states in $\Ket{\Psi}$ due to the quantum parallelism. Besides, the complexity of computing $R(y')$ in step 5(e) is the same as computing $R(y)$, which means
\begin{equation}
        \textrm{TC}_\textrm{5\textrm{(c)}}=\textrm{TC}_\textrm{5\textrm{(e)}}=\textrm{TC}_{5\textrm{(b)}}=O\left(2^tk\right)+O\left(kn\right).
        \label{equa:6}
\end{equation}

To find the index of a marked item among $K$ items where $r-1$ items are marked, the expected number of iterations used by the quantum searching algorithm is at most $(9/2)\sqrt{K/(r-1)}$~\cite{Ref10}; hence, the complexity of step 5(d) is
\begin{equation}
        \textrm{TC}_{5\textrm{(d)}}=O\left(\sum_{r=2}^{K}\frac{1}{r}\cdot\frac{9}{2}\sqrt{\frac{K}{r-1}}\right)=O\left(2^{k/2}\right).
    \label{equa:7}
\end{equation}

Thus, the total computational complexity of our quantum differential attack is
\begin{equation}
    \begin{aligned}
        \textrm{TC}_\textrm{total}&=\textrm{TC}_{(4)}+\textrm{TS}_{5\textrm{(a)}}+\textrm{TS}_{5\textrm{(b)}}+\textrm{TS}_{5\textrm{(c)}}+\textrm{TS}_{5\textrm{(d)}}+\textrm{TS}_{5\textrm{(e)}}\\
                                  &=O\left(2^{k/2}\right)+O\left(2^tk\right)+O\left(kn\right).
    \end{aligned}
    \label{equa:8}
\end{equation}

If we want to estimate the rotation angle $\theta$ caused by each application of the $G$ to $m=\left\lceil n/2 \right \rceil +1$ bits of accuracy with success probability at least $1-\epsilon = 90\%$ when computing $R(\cdot)$, then $t=\left\lceil n/2\right\rceil+4$, $2^t\leq 16\cdot 2^{(n+1)/2}$ and
\begin{equation}
        \textrm{TC}'_\textrm{total}=O\left(2^{k/2}\right)+O\left(k2^{(n+1)/2}\right)+O(kn)=O\left(\sqrt{N}\right)+O\left(\sqrt{K}\right).
    \label{equa:9}
\end{equation}
\subsection{Storage complexity}
\label{sec:4.2}
The memory space needed by our quantum differential cryptanalysis involves two parts: the classical space for $N$ pairs and the quantum space for the registers used in Fig.~\ref{fig:1}. The space cost classically is $O(N)$, and the circuit of the quantum differential attack needs four registers which totally holds $2k+n+t+1$ qubits. If we set $t=\left\lceil n/2 \right\rceil +4$ as before, the quantum storage complexity of this algorithm is $O\left(\log K\right)+O\left(\log N\right)$.
\subsection{Success probability}
\label{sec:4.3}
To ensure that our quantum differential cryptanalysis returns the correct subkey, two conditions must be satisfied: 1. The number of right pairs of the correct subkey estimated by the quantum counting algorithm is more than the estimated number of right pairs of the rest incorrect subkeys; 2. The maximum-finding algorithm described by step (5) can successfully locate the subkey whose estimated number of right pairs is the most.

By the analysis in ~\cite{Ref11}, we know that if the total running time of step (5) is $T=2cm_0$, where $m_0$ is the expected number of time steps of (5), then the second condition can be satisfied with probability at least $1-1/2^c$. If we set $c$ appropriately, this probability could be high enough.

Although the quantum counting algorithm can just estimate the number of solutions approximately, it does not matter too much for our quantum differential cryptanalysis, because the number of the right pairs of any incorrect subkeys is near to zero and much smaller than that of the correct subkey. For example, when attacking DES, the expected number of right pairs of the correct key $z$ and any incorrect key $i$ are $\overline{R}(z)=N\cdot p$ and $\overline{R}(i)=N\cdot p\cdot 2^{-4k/6}$, where $p$ is the probability of the characteristic and $0\leq z,i\leq 2^k-1$. That is, when $k=18$ and $\overline{R}(z)=8$,  the expected number of right pairs of any incorrect subkey is $\overline{R}(i)=1/512$. If we choose $m=\left\lceil n/2 \right\rceil +1$, $\epsilon =1/10$ and thus $t=\lceil n/2 \rceil +4$, the estimation accuracy $|\triangle M|$ of the number of right pairs denoted by $M$ satisfies
\begin{equation}
    |\triangle M|<\left(\sqrt{2MN}+N/2^{m+1}\right)\cdot 2^{-m}\leq \sqrt{M/2}+1/8,
\label{equa:4.15}
\end{equation}
which means the estimated number of right pairs of $z$ and $i$ will be in the range $5+7/8<R(z)<10+1/8$ and $-79/512<R(i)<81/512$ with probability at least $90\%$. It follows that the estimated number of right pairs of incorrect subkeys is close to zero and the quantum differential cryptanalysis is able to find the correct subkey with high probability.
\section{Conclusion}
\label{sec:5}
This paper presents a quantum mechanical differential cryptanalysis method which runs much faster than the classical one.

The heart of differential cryptanalysis (both classical and quantum) is made up of counting procedure and maximum-finding process. Due to the quantum parallelism, only one counting procedure with time $O\left(\sqrt{N}\right)$ is needed to obtain the number of right pairs form $K$ candidate subkeys. Based on these counting results, the quantum maximum-finding process can locate the correct subkey in $O\left(\sqrt{K}\right)$. By contrast, if the memory space for the classical differential cryptanalysis is not large enough, the counting procedure for each candidate subkey need to be carried out one by one and the computational complexity reaches $O\left(KN\right)$, which is impractical in most cases. Even if huge memory is offered so that the well-known counting method\footnote{The counting method in differential cryptanalysis is presented by Biham and Shamir in ~\cite{Ref9}, and it needs huge numbers of counters ($K$ counters are necessary) and many precomputed differential tables.} can be adopted, we still have to traverse $N$ pairs to gain the number of right pairs of the candidate subkeys, and then $K$ steps is needed to get the largest one in the previous outcome.

Hence, the time complexity of this quantum differential attack is $O\left(\sqrt{N}\right)+O\left(\sqrt{K}\right)$, while the best classical differential analysis based on counting procedures needs $O(N)+O(K)$ steps. Moreover, the circuit of our quantum differential cryptanalysis can be modified conveniently to implement any classical attack which is based on counting and maximum-finding procedures, such as the basic linear cryptanalysis ~\cite{Ref13} and the extensions of it.

\begin{acknowledgements}
We gratefully acknowledge the support of the National Natural Science Foundation of China under Grant No. 61173050. The fourth author also gratefully acknowledges the support from the China Postdoctoral Science Foundation under Grant No.
2014M552041.\end{acknowledgements}


\begin{thebibliography}{}
%
%
\bibitem{Ref1}
Shor P W. Algorithms for quantum computation: Discrete logarithm and factoring. In: 35th Annual Symposium on IEEE Foundations of Computer Science, 1994 Proceedings, pp.124--134 (1994)

\bibitem{Ref2}
Boneh D, Lipton R J. Quantum cryptanalysis of hidden linear functions. In: CRYPTO95, pp.424--437. Springer, Berlin(1995)

\bibitem{Ref3}
Grover L K.  Quantum mechanics helps in searching for a needle in a haystack. Phys. Rev. Lett. 79(2), 325--328 (1997)

\bibitem{Ref5}
Ludwig C. A faster Lattice reduction method using quantum search.Springer, Berlin (2003)

\bibitem{Ref6}
Phaneendra H D, Vidya R C, Shivakumar M S. Applying quantum search to a known-plaintext attack on two-key triple encryption. Int. Fed. Inf. Process. 228, 171--178 (2006)

\bibitem{Ref7}
Zhong P C. Bao W S. Quantum mechanical meet-in-the-middle search algorithm for triple-DES. Chin. Sci. Bull. 55(3), 321--325 (2010)


\bibitem{Ref9}
Biham E, Shamir A. Differential cryptanalysis of the data encrypt standard. Springer, New York (1993)

\bibitem{Ref10}
Boyer M, Brassard G, H\'oyer P, et al. Tight bounds on quantum searching (1996). arXiv preprint: arXiv.quant-ph/9605034

\bibitem{Ref11}
Durr C, H\o yer P. A quantum algorithm for finding the minimum (1996). arXiv preprint: arXiv.quant-ph/9607014

\bibitem{Ref12}
Nielsen M A, Chuang I L. Quantum computation and quantum information, 261--263. Cambridge university press, New York (2010)

\bibitem{Ref13}
Matsui M. Linear cryptanalysis method for DES cipher. In: Workshop on the Theory and Application of of Cryptographic Techniques, pp.386--397  Springer, Berlin  (1994)

\bibitem{Ref14}
Biham E, Shamir A. Differential cryptanalysis of the full 16-round DES. In: Differential Cryptanalysis of the Data Encryption Standard, pp. 79--88. Springer Berlin (1993)

\end{thebibliography}


\end{document}